\documentclass[11pt]{revtex4-1}

\usepackage{graphicx}

\begin{document}

\title{A study of different horizons in inhomogeneous LTB cosmological model}

\author{Subenoy Chakraborty\footnote {schakraborty.math@gmail.com}}
\author{Subhajit Saha\footnote {subhajit1729@gmail.com}}
\affiliation{Department of Mathematics, Jadavpur University, Kolkata 700032, West Bengal, India.}

\begin{abstract}
This work deals with a detailed study of the dynamics of the apparent, event and particle horizons in the background of the inhomogeneous LTB spacetime. The comparative study among these horizons shows a distinct character for apparent horizon compared to the other horizons. The apparent horizon will be a trapping horizon if its acceleration is positive. The Kodama vector is also defined and its causal character is found to be similar to that in the FRW model.

\end{abstract}

\maketitle
\section{INTRODUCTION}

In 1970's there was a major advancement in the field of theoretical physics when Hawking, using semi-classical description, showed that a BH behaves as a black body emitting thermal radiation with temperature (known as Hawking temperature) and entropy (known as Bekenstein entropy) proportional to the surface gravity at the horizon and area of the horizon \cite{Hawking1, Bekenstein1} respectively. Further, this Hawking temperature and Bekenstein entropy are related to the mass of the BH through the first law of thermodynamics \cite{Bardeen1}. Thus black hole thermodynamics plays the role of a connecting path between classical gravity and quantum mechanics. Also there was a natural speculation about the inter relationship between the BH thermodynamics and the Einstein field equations as the physical quantities like temperature and entropy on the horizon are associated with the geometry of the horizon. In 1995, Jacobson \cite{Jacobson1} came forward with deriving Einstein field equations from the first law of thermodynamics for all local Rindler causal horizons. Then Padmanabhan \cite{Padmanabhan1, Padmanabhan2} from the other side was able to derive the first law of thermodynamics on the horizon starting from Einstein equations for a general static spherically symmetric spacetime. However, due to nonavailability of a full theory of quantum gravity, it is not possible to formulate a statistical mechanics for microscopic picture of the BH thermodynamics.

On the other hand, this nice idea of equivalence between Einstein field equations and the thermodynamical laws has been extended in the context of cosmology. Gibbons and Hawking \cite{Gibbons1} considered de Sitter cosmological event horizon as a thermodynamical system having temperature and entropy, similar to the Schwarzschild horizon. Normally for an event horizon, one requires to know the entire future development and the causal structure of the spacetime and it does not exist in general for dynamical spacetime due to nonavailability of timelike killing vector. However, static spacetime (having high symmetry) admits a timelike killing vector and event horizon plays a role analogous to that of the Schwarschild event horizon. Usually in dynamical situations, outermost marginally trapped surfaces and apparent horizons are used as proxies for event horizons \cite{Baumgarte1}. As an apparent horizon always exists in a FRW spacetime so for this case universe bounded by apparent horizon is usually considered to be a thermodynamical system having entropy and temperature given by
\begin{equation}
{S_A}=\frac{\pi R_{A}^2}{G}~~and~~{T_A}=\frac{1}{2\pi R_A}
\end{equation}
respectively, where $R_A$ is the radius of the apparent horizon. Recently, using the tunnelling approach \cite{Parikh1} of Parikh and Wilczek and different definitions of surface gravity, different notions of temperature have been manifested of which Kodama-Hayward prescription \cite{Kodama1} seems to be the best. Although there exists no timelike killing vector still Kodama vector \cite{Kodama1} is associated with a conserved current \cite{Abreu1} and the associated Noether charge is nothing but the Misner-Sharp-Hernandez mass \cite{Misner1, Hernandez1}, which can be identified as the internal energy in horizon thermodynamics. Further, in spherically symmetric model using Kodama-Hayward surface gravity to thermodynamics, it is possible to show equivalence of the Misner-Sharp-Hernandez mass to Hawking quasilocal energy \cite{Hayward1}.

In the present work, we consider the inhomogeneous LTB spacetime model and examine the dynamics of the horizons namely particle, event and apparent horizons with the evolution of the universe. The choice of the inhomogeneous LTB model of the universe is as follows: Based on the form of the luminosity distance as a function of the redshift for distant supernova \cite{Riess1, Riess2, Perlmutter1, Percival1, Sievers1}, it is speculated that our universe is going through an accelerated expansion in the present epoch. The observed perturbations in the cosmic microwave background \cite{Spergel1} supports this recent accelerating phase for homogeneous cosmology although no definite mechanism is known so far which is responsible for this late time acceleration. The simplest mechanism that one can think of is the inclusion of a non-zero cosmological constant. However, the value of the cosmological constant estimated from the fact that there is no acceleration beyond $z\sim 1$ (i.e., $z\gtrsim 1$), is approximately $120$ orders of magnitude smaller than its natural value in Planck scale.

On the other hand, recently it has been claimed that local inhomogeneities, on account of their backreaction on the metric, are responsible for the apparent acceleration of the universe \cite{Pascual-Sanchez1, Rasanen1}. Thus accelerated expansion is not necessarily required to fit the data provided inhomogeneous models of the universe are considered. The simplest inhomogeneous toy model corresponding to spherical symmetry is the Lemaitre-Tolman-Bondi (LTB) model \cite{Lemaitre1, Lemaitre2, Tolman1, Bondi1} of the universe. Also LTB model serves as a simple testing ground for the effects of inhomogeneities when cosmological data are fitted without dark energy. This paper is organized as follows: Sec. II deals with the basic equations related to LTB model. A comparative study of the three horizons (namely particle, event and apparent) has been done extensively in Sec. III. The Kodama vector has been defined and its causal character has been investigated in Sec. IV. Finally, a brief discussion and concluding remarks has been presented in Sec. V.
 
\section{THE LTB MODEL}

The line element for LTB model in comoving coordinates ($t$, $r$, $\theta$, $\phi$) is given by
\begin{equation}
ds^2=-dt^2+\frac{{R'}^2}{1+f(r)}dr^2+R^2(d{\theta}^2+sin^2\theta d{\phi}^2),
\end{equation}
where $(d{\theta}^2+sin^2\theta d{\phi}^2)=d\Omega_{2}^{2}$ is the metric on a unit two-sphere, $R=R(r,t)$ is the area-radius of the spherical surface, $f(r)$ is the curvature scalar which classifies the spacetime as bounded, marginally bounded or unbounded according as $f(r)<0$, $f(r)=0$ or $f(r)>0$ and overdash and dot indicate differentiation with respect to $r$ and $t$ respectively.\\\\
In general relativity, the Einstein equations for LTB model of the dusty universe are written as (choosing $8\pi G=1=c$)
\begin{equation}
\frac{F'(r)}{R^2 R'}=\rho
\end{equation}
and the evolution equation for $R$ is given by
\begin{equation}
\dot{R}^2=\frac{F(r)}{R}+f(r).
\end{equation}
The covariant conservation equation ${T_{\nu}^{\mu}}_{;\mu}=0$ gives the energy conservation equations
\begin{equation}
\dot{\rho}+3H\rho=0.
\end{equation}
In the above Einstein equations $F(r)=R({\dot{R}}^2-f(r))$ is termed as the mass function, $\rho$ is the energy density and $H=\frac{1}{3}\left(2\frac{\dot{R}}{R}+\frac{\dot{R}'}{R'}\right)$ is the average Hubble parameter (which will be discussed later). Here fluid may be considered as succesive shells levelled by $r$, having local density $\rho$ as time dependent. Also $R(r, t)$ stands for location of the shell marked by $r$ at time $t$ having initial condition (by proper rescaling) \cite{Joshi1}
\begin{equation}
R(r, 0)=r.
\end{equation}

In LTB model, the proper volume of a sphere of radius $R$ at $t=constant$ hypersurface can be written in comoving coordinates as
\begin{equation}
V_P=\int_0^{2\pi} d\phi \int_0^\pi d\theta \int_0^rdr' \sqrt{^{(3)}g},
\end{equation}
with ${^{(3)}g}=\frac{R'^2 R^4}{1+f(r)}sin^2\theta$, the determinant of the metric in the 3D hypersurface $t=constant$. If we introduce the hyperspherical radius $\chi$ and choose the curvature scalar $f(r)$ as $\epsilon R^2$ ($\epsilon =+1$ for unbounded space, $\epsilon =-1$ for bounded space and $\epsilon =0$ for marginally bounded space), then the proper volume has the explicit form
\begin{eqnarray}
V_P=2\pi \left[R\sqrt{1+R^2}-sinh^{-1} R \right]~,~~\epsilon =+1\\
V_P=\frac{4\pi}{3}R^3~,~~\epsilon =0\\
V_P=2\pi \left[sin^{-1}R-R\sqrt{1-R^2} \right]~,~~\epsilon =-1,
\end{eqnarray}
so that only in the flat case, $\rho V_P$ is the Misner-Sharp-Hernandez mass.\\\\
Now the volume expansion rate is defined in terms of the four velocity of the fluid $u^\alpha$ as 
\begin{equation}
\theta=3H={u^{\alpha}}_{;\alpha}=u_{\alpha;\beta}g^{\alpha \beta}=u_{\alpha,\beta}h^{\alpha \beta},
\end{equation}
with $h^{\alpha \beta}=g^{\alpha \beta}+u^{\alpha}u^{\beta}$ be the projection tensor in 3-space. The Roychoudhuri equation for this inhomogeneous spacetime takes the form
\begin{equation}
\dot{H}\equiv H_{,\alpha}u^{\alpha}=-H^2-\frac{1}{3}\sigma_{ab}\sigma^{ab}+\frac{1}{3}\omega_{ab}\omega^{ab}-\frac{1}{3}R_{ab}u^{\alpha}u^{\beta},
\end{equation}
where $\sigma_{ab}$ and $\omega_{ab}$ are the shear and vorticity tensors respectively. So for pressureless (i.e., dust) irrotational fluid, the deceleration parameter (apparent acceleration) of a comoving observer is given by
\begin{equation}
q_A=-\left(1+\frac{\dot{H}}{H^2}\right)=\frac{1}{H^2}\left(\frac{1}{3}\sigma_{ab}\sigma^{ab}+\frac{\rho}{12M^2}\right).
\end{equation}
Thus if the local energy density is positive then there will always be deceleration.\\\\
It should be noted that due to inhomogeneity of the spacetime the above definition of expansion rate does not truly reflect the variation of the expansion rate in different directions. In particular, in the present LTB cosmological model, radial direction is a preferred direction, so one can define a tensor $P^{ab}$ that projects every quantity perpendicular to the preferred spacelike direction ($s^a$) as \cite{Apostolopouls1}
\begin{equation}
P^{ab}=g^{ab}+u^a u^b-s^a s^b=h^{ab}-s^a s^b.
\end{equation}
For the present model, $s^a\equiv\frac{\sqrt{1+f(r)}}{R'}\frac{\partial}{\partial r}$. Thus we can define invariant expansion rates parallel and perpendicular to the preferred direction (i.e., $s^a$) as
\begin{equation}
H_r=u_{a;b}s^a s^b=\frac{\dot{R}'}{R'}
\end{equation}
and
\begin{equation}
H_\Omega=\frac{1}{2}u_{a;b}P^{ab}=\frac{\dot{R}}{R},
\end{equation}
so that the average of the expansion rates in various directions is given by 
\begin{equation}
H=\frac{1}{3}(2H_\Omega +H_r).
\end{equation}
Moreover, using this expansion rate $H$ we have seen that the expansion is always decelerating. However, it may so happen that the expansion is accelerating in some direction but still the average expansion rate is decelerating. Hence to take account of the directional dependence we define the generalized expansion rate as \cite{Partovi1, Humphreys1}
\begin{equation}
H_G=\frac{1}{3}{u^a}_{;a}+\sigma_{ab}e^a e^b,
\end{equation}
where $e^a$ is the unit vector along the direction of observation and $H_G$ stands for generalized Hubble parameter. For the present model, the explicit value of $H_G$ is \cite{Partovi1}
\begin{equation}
H_G=\frac{\dot{R}}{R}+\left(\frac{\dot{R}'}{R'}-\frac{\dot{R}}{R}\right)cos^2 \psi,
\end{equation}
where $\psi$ is the angle between the radial direction through the observer and the direction of observation. Hence we have, $H_G=H_r$ for $\psi =0$ or $\pi$ and $H_G=H_\Omega$ for $\psi =\frac{\pi}{2}$ or $\frac{3\pi}{2}$. Then one can define the deceleration parameter in a specific direction as (for small $z$) \cite{Apostolopouls1, Partovi1}
\begin{equation}
q_G=-H_G\frac{d^2 D_L}{dz^2}+1,
\end{equation}
where $D_L$ stands for the expansion of the luminosity distance of a light source in powers of the redshift $z$ of the incoming photons. Hence the deceleration parameter along and perpendicular to the radial direction are \cite{Apostolopouls1}
\begin{equation} \label{qr}
q_r={\frac{R'}{\dot{R}'}}^2 \left[\frac{\sqrt{1+f(r)}}{R'}\frac{\partial}{\partial r} \left(\frac{\dot{R}'}{R'}\right)-\left(\frac{\ddot{R}'}{R'}\right)\right]
\end{equation}
and
\begin{equation} \label{qo}
q_\Omega=-\left(\frac{R}{\dot{R}}\right)^2 \frac{\ddot{R}}{R}.
\end{equation}
Now due to positive indefiniteness of the expressions on the right hand side of Eqs. (\ref{qr}) and (\ref{qo}), $q_r$ and $q_\Omega$ may have sign difference at any point of the LTB spacetime, i.e., at any point it is possible to have acceleration along the radial direction and deceleration along the cross radial direction and vice-versa.

\section{HORIZONS IN LTB MODEL}

\subsection{Particle horizon and event horizon: A comparative study}

We shall now discuss the two familiar horizons in cosmology, namely the particle horizon and the event horizon. Let us consider the world line of an observer $O$ moving on a timelike geodesic in a spacetime for which $I^-$ is spacelike. The past lightcone at any point $P$ on the world line consists of events in spacetime which can be observed by $O$ at that time. The division of particles into those seen by $O$ at $P$ and those not seen by $O$ at $P$ gives rise to the notion of particle horizon of $O$ at $P$. So particle horizon is nothing but the limit of $O$'s vision. Hence the particle horizon at any time $t$ for a comoving observer at $r=0$ is the spherical surface having centre at $r=0$ and radius (considering radial null geodesics $r=r(t)$ for the LTB metric (2))
\begin{equation}
R_{PH}(t,r)=\frac{R'(t,r)}{\sqrt{1+f(r)}}\int_0^t \frac{\sqrt{1+f(r)}}{R'(t',r)}dt^{'}.
\end{equation}
If this integral is a proper integral (or convergent improper integral) then the observer at $r=0$ receives at time $t$ only those light signals which started within the sphere of proper radius $R_{PH}$. On the other hand, if the above improper integral diverges then the maximal volume that can be causally connected to the observer at time $t$ is infinite, i.e., he receives all the light signals emitted at any early time of the evolution of the universe. We say that particle horizon does not exist in this case. This is the situation when $I^-$ is null.\\\\
The velocity and the acceleration of the particle horizon are given by
\begin{equation}
\dot{R}_{PH}=\left\lbrace H_r-\frac{f^{'}(r)}{2R^{'}\sqrt{1+f(r)}}\right\rbrace R_{PH}+1
\end{equation}
and
\begin{equation}
\ddot{R}_{PH}=\left\lbrace \dot{H}_r+H_r^2-\frac{{H_r}f^{'}(r)}{2R^{'}\sqrt{1+f(r)}}-\frac{f^{''}(r)}{2{R^{'}}^2}+\frac{\lbrace{f^{'}(r)\rbrace}^2}{2{R^{'}}^2\lbrace{1+f(r)\rbrace}}\right\rbrace R_{PH}+\left\lbrace H_r-\frac{f^{'}(r)}{2R^{'}\sqrt{1+f(r)}}\right\rbrace .
\end{equation}
respectively with $H_r=\frac{{\dot{R}}'}{R'}$, the radial Hubble parameter. Note that in the above equations, radial derivative acting on $f$ appears as we are considering radial null geodesics $r=r(t)$. However, for marginally bound case (i.e., $f(r)=0$), there is no longer any radial derivative and velocity and acceleration of the particle horizon have the simple expressions
\begin{equation}
\dot{R}_{PH}={H_r} R_{PH}+1
\end{equation}
and
\begin{equation}
\ddot{R}_{PH}=(\dot{H}_r+H_{r}^{2})R_{PH}+H_r.
\end{equation}
Note that the particle horizon is commonly studied in inflationary cosmology. In an expanding universe the radius of the particle horizon (if exists) always increases and as a result more and more light signals emitted between the big bang and time $t$ will be responded by the observer at $O$. However, for finite $R_{PH}$, there will always be a region inaccessible to the comoving observers \cite{Faraoni1}. Usually the particle horizon is thought of as a timelike surface, i.e., the boundary of the evolution of all particles in the causal past of an observer.\\

On the other hand, if both $I^-$ and $I^+$ are spacelike in a spacetime and if we consider the whole history of the observer $O$, then the past light cone of $O$ at $P$ on $I^+$ is called the future event horizon of $O$. So events outside this horizon will never be seen by $O$. However, if $I^+$ is null and the observer $O$ moves on a timelike geodesic, then there is no longer any event horizon. But for an uniform accelerated observer, the world line ends up on $I^+$ and future event horizon exists. Thus these event horizons are observer dependent in contrast to BH event horizon (which is absolute in nature). Thus, event horizon is the boundary of the spacetime region and which consists of all the events between the present time $t$ and the future infinity $t=\infty$ which can be accessible to a comoving observer at $O$. The proper radius of the event horizon for the  present LTB model is given by
\begin{equation}
R_{EH}=\frac{R'(t,r)}{\sqrt{1+f(r)}}\int_t^\infty \frac{\sqrt{1+f(r)}}{R'(t',r)}dt^{'},
\end{equation}
and one may say that $R_{EH}$ is the proper distance to the most distant event the observer will ever see.
As before, if the above improper integral converges (i.e., $R_{EH}$ will have a finite value) then the observer at $O$ will never be aware of events beyond $R_{EH}$, while divergence of this improper integral means that the observer can see events arbitrarily far away from him (may be at sufficiently large time).\\

From the definition of the above two horizons, one might say that event horizon is the complement of the particle horizon. Further, from the very definition of the event horizon, we see that a knowledge of the entire future history of the universe starting from the present epoch is necessary. In fact, the event horizon is the boundary of the past of future null infinity and is global in nature. Moreover, comparing to the blackhole event horizon, we have the following basic differences:\\
$\bullet$ Both the cosmological horizons (particle and event) are observer dependent while BH event horizon is an universal one.\\
$\bullet$ In case of cosmological event horizon, the observer is located inside it and so he is not able to communicate with events outside event horizon, while the observer is always situated outside the BH horizon and he is not able to interact with events inside the BH horizon.\\

As the cosmological event horizon is a causal boundary, so it is a null surface generated by null geodesics. In comoving coordinates, the surface of the event horizon is characterized by
\begin{equation}
\frac{dt}{R^{'}}-\frac{dr}{\sqrt{1+f(r)}}=0.
\end{equation}\\
So the evolution of the event horizon is described by its velocity and acceleration as
\begin{equation}
\dot{R}_{EH}=\left\lbrace H_r-\frac{f^{'}(r)}{2R^{'}\sqrt{1+f(r)}}\right\rbrace R_{EH}-1
\end{equation}
and
\begin{equation}
\ddot{R}_{EH}=\left\lbrace \dot{H}_r+H_r^2-\frac{{H_r}f^{'}(r)}{2R^{'}\sqrt{1+f(r)}}-\frac{f^{''}(r)}{2{R^{'}}^2}+\frac{\lbrace{f^{'}(r)\rbrace}^2}{2{R^{'}}^2\lbrace{1+f(r)\rbrace}}\right\rbrace R_{EH}-\left\lbrace H_r-\frac{f^{'}(r)}{2R^{'}\sqrt{1+f(r)}}\right\rbrace .
\end{equation}
which have the simple expressions
\begin{equation}
\dot{R}_{EH}={H_r}R_{EH}-1
\end{equation}
and
\begin{equation}
\ddot{R}_{EH}=(\dot{H}_r+H_r^{2})R_{EH}-H_r
\end{equation}
for marginally bound case (i.e., $f(r)=0$).\\

For the present inhomogeneous model, we are not able to infer about the existence of event (or particle) horizon as it can be done for homogeneous FRW model (event horizon exists only for accelerating phase). However, from the mathematical point of view, the integral for event (particle) horizon is an improper integral of the first (second) kind and its convergence is the criteria for the existence of the event horizon. Due to the complexity of the model, we choose (as a toy model) a simple form of the scale factor $R(r,t)$ as 
\begin{equation}
R(r,t)=\chi (r) t^\alpha ~,~~~\alpha~constant,
\end{equation} 
(Note that the choice $R(r,t)=r\xi (t)$ results the FRW model) where $\chi (r)$ is assumed to be at least twice differentiable. Then we have(on choosing $f(r)=\lbrace{{\chi ^{'}(r)}\rbrace}^2-1$),
\begin{equation}
R_E=\frac{t}{\alpha -1},
\end{equation}
and it exists only for $\alpha >1$, while
\begin{equation}
R_{PH}=\frac{t}{1-\alpha},
\end{equation}
and it exists only for $\alpha <1$.\\

For $\alpha =1$, both the horizons do not exist. So for this simple choice, both the horizons do not exist simultaneously. But it is not a general feature of the model because if the functional form of $R$ be such that it is finite both at $t=0$ and at $t=\infty$ then both the horizons might exist simultaneously. Further, for this typical choice of $R$, the deceleration parameter along and perpendicular to the radial direction of an observer are
\begin{equation}
q_r=-\frac{\alpha -1}{\alpha}=q_\Omega,
\end{equation} 
i.e., the acceleration is isotropic for this simple choice. Also for this simple choice, the existence (or non-existence) of event horizon is related to the acceleration (or deceleration) of the universe.

\subsection{Apparent horizon}

Analogous to the FRW spacetime, we consider hypersurfaces characterized by constant comoving time and the comoving components of the tangent fields corresponding to the outgoing and ingoing radial null geodesics are given by
\begin{equation} \label{nmu}
l^\mu =\left(1, \frac{\sqrt{1+f(r)}}{R'}, 0, 0 \right)~~and~~n^\mu =\left(-1, \frac{\sqrt{1+f(r)}}{R'}, 0, 0 \right)
\end{equation}  
respectively with $l^\mu l_\mu =0=n^\mu n_\mu$ and $l^\mu n_\mu =2$. Note that we have freedom to rescale null vectors and the common normalization is chosen such that $l^\mu n_\mu =-1$ (i.e., dividing both the null vectors by a factor of $\frac{1}{\sqrt{2}}$). Using the standard expression for the expansion scalar corresponding to a null geodesic congruence namely
\begin{equation}
\theta _l=\left[g^{\mu \nu}+\frac{l^\mu n^\nu +n^\mu l^\nu}{(-g_{\alpha \beta}n^\alpha l^\beta)}\right]{\nabla}_\mu l_\nu,
\end{equation}
we obtain
\begin{equation}
\theta _l=2\left(H_\Omega +\frac{\sqrt{1+f(r)}}{R}\right)~~and~~\theta _n=2\left(H_\Omega -\frac{\sqrt{1+f(r)}}{R}\right),
\end{equation}
where $H_\Omega =\frac{\dot{R}}{R}$ is the angular Hubble parameter.\\

The apparent horizon is usually defined as the outermost marginally outer trapped surface. Mathematically, the apparent horizon is a surface defined by the conditions on the expansion scalars for null geodesic congruences (outgoing and ingoing) as
\begin{equation}
\theta _l>0~,~~\theta _n=0,
\end{equation}
which gives
\begin{equation} \label{rhh}
R_{AH}=\frac{\sqrt{1+f(r)}}{H_\Omega}=R_{HH}\sqrt{1+f(r)}.
\end{equation}
Here, in analogy to the FRW model, $R_{HH}=\frac{1}{H_\Omega}$ is referred to as the Hubble horizon. 
Note that as apparent horizon is characterized locally by the null geodesic congruences and their expansions, so it is not related to the global causal structure.\\\\
Further, 
\begin{equation}
\theta _l \theta _n=\frac{4(1+f(r))}{R^2}\left(\frac{R^2}{R_{AH}^2}-1\right).
\end{equation}\\
Thus we have $\theta _l>0$~,~~$\theta _n>0$ in $R>R_{AH}$ and $\theta _l>0$~,~~$\theta _n<0$ in $0 \leq R<R_{AH}$.
Thus radial null rays from the region outside the apparent horizon fails to cross the horizon and reach the observer \cite{Faraoni1}.\\

From Eq. (\ref{rhh}), we see that the apparent horizon coincides with the Hubble horizon in the marginally bound case (i.e., $f(r)=0$). The apparent horizon is covered by the Hubble horizon or it covers the Hubble horizon for bounded or unbounded LTB model respectively.\\

The surface of the apparent horizon in comoving coordinates can be described as
\begin{equation}
\Phi (r,t)\equiv R(r,t)-F(r)=0.
\end{equation} \\
The velocity and the acceleration of the apparent horizon characterizing its evolution are given by
\begin{equation}
\dot{R}_{AH}=-\frac{\dot{H}_{\Omega}}{H_{\Omega}}R_{AH}+\frac{f^{'}(r)}{2R^{'}H_{\Omega}}
\end{equation}
and
\begin{equation}
\ddot{R}_{AH}=-\left\lbrace \frac{\ddot{H}_{\Omega}}{H_{\Omega}}-2\frac{\dot{H}_{\Omega}^{2}}{H_{\Omega}^{2}}\right\rbrace R_{AH}-\left\lbrace 2\frac{\dot{H}_{\Omega}}{H_{\Omega}^{2}}+\frac{H_r}{H_{\Omega}}\right\rbrace \frac{f^{'}(r)}{2R^{'}}+\frac{f^{''}(r)\sqrt{1+f(r)}}{2{R^{'}}^2 H_{\Omega}}.
\end{equation}\\
The normal to this surface is given by
\begin{eqnarray}
n_\mu = {\nabla _\mu \Phi |_{AH}} &=& \lbrace{(\dot{R}-\dot{F})\delta _{\mu 0}+(R'-F')\delta _{\mu 1}\rbrace}|_{AH} \nonumber \\
&=& -2H_{\Omega}R_{AH}^{2}\ddot{R}_{AH} \delta _{\mu 0}-R_{AH}^{2}R_{AH}^{'}\left\lbrace 2{H_{\Omega}}H_r-\frac{f^{'}(r)}{R_{AH}R_{AH}^{'}}\right\rbrace \delta _{\mu 1},
\end{eqnarray}
with norm squared
\begin{equation}
{\parallel n_\mu \parallel}^2=-R^4 \left[4H_{\Omega}^2\ddot{R}_{AH}^2-{R_{AH}^{'2}}\left\lbrace 2{H_\Omega}H_r-\frac{f^{'}(r)}{R_{AH}R_{AH}^{'}}\right\rbrace ^2 \right].
\end{equation}
In particular, for marginally bound case (i.e., $f(r)=0$), we have
\begin{equation}
{\parallel n_\mu \parallel}^2=-4R^4{H_{\Omega}}^2\lbrace{\ddot{R}^2-{{\dot{R}}^{'2}}\rbrace}.
\end{equation}
Hence from the above expression for the (norm) normal vector, we are not able to draw any definite conclusion about the causal character of the apparent horizon (i.e., null, timelike or spacelike).\\

Finally, to have a comparative study of the evolution of the horizons, we first note that normally all of them are not comoving. The difference between the expansion rate of a horizon and that of the comoving radius is given by
\begin{equation}
\frac{\dot{R}_{PH}}{R_{PH}}-H_\Omega =\Delta +\frac{1}{R_{PH}}-\frac{f^{'}(r)}{2R_{PH}^{'}\sqrt{1+f(r)}},
\end{equation}
\begin{equation}
\frac{\dot{R}_{EH}}{R_{EH}}-H_\Omega =\Delta -\frac{1}{R_{EH}}-\frac{f^{'}(r)}{2R_{EH}^{'}\sqrt{1+f(r)}}
\end{equation}
and
\begin{equation}
\frac{\dot{R}_{AH}}{R_{AH}}-H_\Omega =\Delta -\left\lbrace \frac{\dot{H_{\Omega}}}{H_{\Omega}}+H_r\right\rbrace +\frac{f^{'}(r)}{2R_{AH}R_{AH}^{'}H_{\Omega}},
\end{equation}
where $\Delta =H_r-H_\Omega =\frac{\dot{R}'}{R'}-\frac{\dot{R}}{R}$ is termed as a measure of inhomogeneity. For marginally bound case, i.e., ($f(r)=0$), as long as $\Delta >0$ or $|\Delta|<\frac{1}{R_{PH}}$, the particle horizon always expands faster than comoving while if $\Delta <0$ or $|\Delta|<\frac{1}{R_{EH}}$, the event horizon expands slower than comoving and it will be comoving if $\Delta =\frac{1}{R_{EH}}$. The apparent horizon might expand faster or slower than comoving.\\

Lastly, it is legitimate to address the question whether the apparent horizon in LTB model is a trapping horizon or not. According to Hayward \cite{Hayward2} (in coordinate invariant criteria) the apparent horizon will be a trapping horizon if
\begin{equation}
L_l \theta _n |_{app.}>0,
\end{equation}
where $L_l$ stands for the lie derivative along the outgoing null geodesic and $\theta _n$ stands for expansion scalar corresponding to incoming null geodesic $n^\mu$ (given in Eq. (\ref{nmu})). In the present model, we have,
\begin{equation}
L_l \theta _n |_{app.}=\frac{\ddot{R}_{AH}}{R_{AH}},
\end{equation}
and hence it is clear that the apparent horizon will be a trapping horizon only if the acceleration of the horizon is positive.

\section{Kodama vector}

We shall now introduce the notion of Kodama vector. If the spacetime is static then Kodama vector $K^\mu$ is in general parallel to the time-like killing vector and consequently defines a preferred class of observers with four velocity $u^\mu =\frac{K^\mu}{\sqrt{|K^\lambda K_\lambda|}}$. 
For a general spherically symmetric metric
\begin{equation}
ds^2=h_{ab}dx^a dx^b+R^2(r,t)d\Omega _{2}^{2},
\end{equation}
the Kodama vector is defined as
\begin{center}
$K^a=\epsilon ^{ab}{\nabla}_b R$,~~~$a$, $b=0$, $1$, 
\end{center}
where $\epsilon ^{ab}$ is the volume form associated with the two-metric and $K^\theta =K^\phi =0$. Note that the Kodama vector is always orthogonal to the two-spheres of symmetry, i.e.,
\begin{center}
$K^a{\nabla}_a R=\epsilon ^{ab}{\nabla}_a R {\nabla}_b R=0$.
\end{center}
Further, it can be shown \cite{Kodama1, Abreu1} that the Kodama vector is divergence free, i.e., $\nabla _a K^a =0$ and consequently the Kodama energy current $J^a=G^{ab}K_b$ is covariantly conserved, i.e., $\nabla ^a J_a=0$ (known as Kodama miracle).\\\\
In comoving coordinates, the explicit form of the Kodama vector is 
\begin{equation}
K^\mu =\left(\sqrt{1+f(r)}, -\sqrt{1+f(r)}\frac{\dot{R}}{R'}, 0, 0 \right)
\end{equation}
and its norm squared is
\begin{equation}
{\parallel K^\mu \parallel}^2=K^\mu K_\mu =(1+f(r))\left\lbrace \frac{R^2}{R_{AH}^{2}}-1 \right\rbrace .
\end{equation}
Thus the Kodama vector is spacelike (i.e., ${\parallel K^\mu \parallel}^2>0$) outside the apparent horizon, timelike (i.e., ${\parallel K^\mu \parallel}^2<0$) inside the apparent horizon and is null (i.e., ${\parallel K^\mu \parallel}^2=0$) on the apparent horizon. As expected, the nature of the Kodama vector is similar to that for the FRW model \cite{Faraoni1} which is a sub-class of LTB model.


\section{SHORT DISCUSSION AND CONCLUDING REMARKS}

The paper begins with an exhaustive study of the LTB model of the universe. Due to inhomogeneity, the Hubble parameter and the deceleration parameter are defined in a generalized way and it is found that though there is an overall deceleration but still there might be an acceleration in some direction. 
The particle and the event horizons are defined by improper integrals but in LTB model, we do not have any physical or mathematical restrictions so that the radius of the event (or particle) horizon is finite. However, assuming the convergence of these improper integrals, velocity and acceleration of the horizons have been evaluated. We have also cited a simple example to show the existence of the horizons and evaluated the deceleration parameter in different directions. Further, using the measure of inhomogeneity, a comparative study of the horizons with comoving radius has been shown and due to complicated expression, one is not able to infer anything about the apparent horizon. Further, it has been shown that depending on the sign of the acceleration of it, the apparent horizon might be a trapping horizon. Finally, the Kodama vector is defined in LTB model and it is found that the Kodama vector is timelike inside the apparent horizon, spacelike outside the apparent horizon and null on the surface of the apparent horizon. Also for the present model, the nature of the Kodama vector inside and outside the apparent horizon is analogous to that in the FRW model.

\begin{acknowledgments}

The authors are thankful to IUCAA, Pune for their warm hospitality as a major part of the work has been done there. Also SC acknowledges the UGC-DRS Programme in the Department of Mathematics, Jadavpur University. The author SS is thankful to UGC-BSR Programme of Jadavpur University for awarding JRF. Both the authors are thankful to the anonymous referee for his valuable comments and suggestions.

\end{acknowledgments}

\end{document}